\documentclass[sigconf]{acmart}
\usepackage{enumitem}
\usepackage{caption}
\usepackage{listings}
\usepackage{todonotes}
\usepackage{xcolor}
\usepackage{multirow}
\usepackage{subcaption}

\AtBeginDocument{%
  }

\copyrightyear{2026}
\acmYear{2026}
\setcopyright{cc}
\setcctype{by}
\acmConference[HPDC '26]{The 35th International Symposium on High-Performance Parallel and Distributed Computing}{July 13--16, 2026}{Cleveland, OH, USA}
\acmBooktitle{The 35th International Symposium on High-Performance Parallel and Distributed Computing (HPDC '26), July 13--16, 2026, Cleveland, OH, USA}
\acmDOI{10.1145/3806645.3807576}
\acmISBN{979-8-4007-2640-8/2026/07}

\begin{document}

\title{{GICC}: A High-Performance Runtime for \underline{G}PU-\underline{I}nitiated \underline{C}ommunication and \underline{C}oordination in Modern HPC Systems}

\author{Baodi Shan}
\email{baodi.shan@stonybrook.edu}
\affiliation{%
  \institution{Stony Brook University}
  \streetaddress{100 Nicolls Road}
  \city{Stony Brook}
  \state{New York}
  \country{USA}
}

\author{Mauricio Araya-Polo}
\affiliation{%
  \institution{TotalEnergies EP Research \& Technology US}
  \streetaddress{P.O. Box 1212}
  \city{Houston}
  \state{Texas}
  \country{USA}
}

\author{Barbara Chapman}
\email{barbara.chapman@stonybrook.edu}
\affiliation{%
  \institution{Stony Brook University}
  \streetaddress{100 Nicolls Road}
  \city{Stony Brook}
  \state{New York}
  \country{USA}
}

\renewcommand{\shortauthors}{Shan et al.}

\begin{abstract}

An increasingly popular approach for distributed GPU applications is kernel-level, cross-node coordination to reduce launch overheads and improve compute--communication overlap. The reality is that such support is lacking. On one hand, on OFI-based interconnects such as HPE Slingshot---which powers six of the top ten systems in the November 2025 Top500, including the top three---GPU kernels cannot autonomously drive distributed coordination: existing runtimes rely on host-driven progress and do not provide a bounded mechanism for recycling pre-staged NIC work across repeated GPU-triggered operations. On the other hand, on InfiniBand, GPU-initiated communication is possible, but current implementations incur unnecessary synchronization and locking overheads.
This paper presents GICC, a GPU-driven coordination framework that enables GPU kernels to issue coordination commands and directly trigger NIC-level operations without host involvement on the fast path. 
For example, in stencil computations, GPU threads can directly initiate halo exchanges with neighboring nodes as soon as boundary regions are computed, rather than waiting for the kernel to complete and relying on the host to orchestrate communication. This eliminates synchronization overhead and enables fine-grained overlap between interior computation and boundary data transfer.
GICC decouples coordination semantics from data movement and introduces an asynchronous resource reclamation scheme: the NIC signals completion to both GPU and host memory, enabling a lightweight host thread to recycle NIC resources concurrently with GPU execution---without injecting latency into the coordination path. This design enables sustained GPU-driven coordination under finite NIC state, a capability absent from existing runtimes on OFI-based fabrics.

We implement GICC on NVIDIA and AMD GPUs over both InfiniBand and Slingshot (OFI). On Slingshot, GICC reduces average per-coordination latency by up to $229\times$ and improves weak scaling efficiency by up to 25\%. On InfiniBand, GICC achieves up to $1.95\times$ lower put latency than NVSHMEM by eliminating unnecessary locking and synchronization. For an industrial stencil-based proxy application on 64 AMD MI250X GCDs, GPU-aware MPI incurs over 52\% higher communication time than GICC, and GICC achieves 42\% parallel efficiency versus MPI's 35.4\%. These results demonstrate that GPU-driven, resource-aware coordination is both necessary and achievable across the dominant interconnects in modern HPC systems.

\end{abstract}



\maketitle
\section{Introduction}

Modern large-scale high-performance computing (HPC) applications increasingly rely on GPUs to deliver extreme computational throughput. In many distributed GPU workloads, execution proceeds as a sequence of short GPU phases---e.g., pipelined stages, fused kernels, or iterative loops---interleaved with coordination points such as barriers, reductions, and phase ordering.

Under today's programming and runtime models, these coordination points are typically host-driven: GPU kernels must return control to the CPU, which invokes a runtime to orchestrate collective operations, enforce ordering, and manage progress. Even when the coordination payload is small, this host round-trip fragments GPU execution, introduces repeated kernel launches and host--device synchronization, and limits the extent to which GPUs can govern distributed control flow. As shown in Section~\ref{sec:problem}, a 200-phase workload can spend over 32\% of its runtime in coordination overhead alone.

\begin{figure}[t]
  \centering
  \includegraphics[width=.95\linewidth]{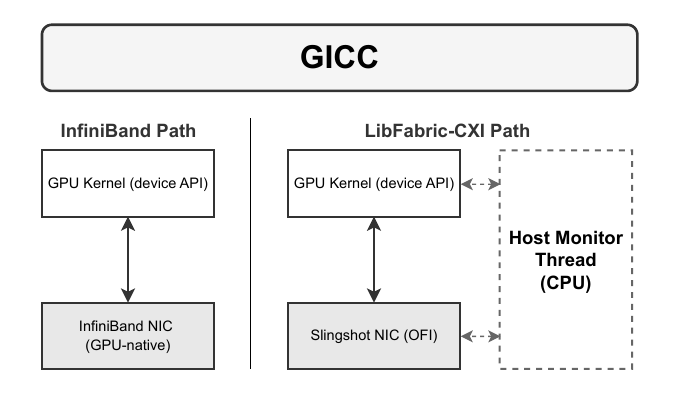}
  \caption{Overview of GICC. On InfiniBand, GPUs directly interact with the NIC on the fast path. On OFI/CXI, GPUs drive coordination within kernels while a lightweight host monitor runs concurrently to provide progress and manage NIC state.}
  \Description{Block diagram of GICC with two fabric-specific paths. The InfiniBand path shows a GPU kernel and an InfiniBand NIC with direct interaction. The OFI/CXI path shows a GPU kernel and a Slingshot/CXI NIC, plus a host monitor thread that runs concurrently and interacts with both the GPU and NIC to provide progress and manage NIC state.}
  \label{fig:gicc_overview}
\end{figure}

GICC addresses two complementary gaps in existing GPU communication frameworks. The first is \emph{GPU-triggered communication} on OFI-based fabrics such as HPE Slingshot. Existing runtimes on Slingshot including NVSHMEM~\cite{nvshmem} are host-mediated---GPU threads may invoke GPU-visible APIs, but the host CPU still mediates work submission and progress. GICC enables GPU-triggered data movement on these fabrics, in which GPU kernels can trigger pre-configured NIC operations without a CPU round trip on the fast path.
The second is \emph{GPU-driven coordination}. Data movement alone is insufficient for distributed GPU applications: while put/get-style operations move bytes, coordination defines ordering and synchronization semantics (e.g., barriers and phase transitions) that must make forward progress under system constraints. We use \emph{GPU-driven coordination} to mean that GPU control flow governs \emph{when} coordination occurs and can observe completion conditions within a kernel. On OFI-based systems, where device code cannot enqueue new network work descriptors, this requires a \emph{GPU-triggered} execution style in which the host pre-configures bounded NIC work and the GPU triggers execution on the fast path. GICC introduces GPU-driven coordination primitives---including active messages and barriers---that allow GPU control flow to govern distributed synchronization across both InfiniBand and OFI/CXI fabrics. Figure~\ref{fig:gicc_overview} summarizes GICC's two-layer design.

Realizing GPU-driven coordination is fundamentally non-trivial. GPU kernels cannot block or issue system calls, lack a general-purpose progress engine, and execute under a SIMD-style model that complicates fine-grained control flow. On the dominant OFI-based interconnects accessed via libfabric~\cite{libfabric}, progress and resource retirement may be mediated by host-side software (e.g., manual provider progress), and modern NICs expose limited on-device state for triggered work. These constraints mean that naive approaches (e.g., simply polling GPU-visible flags or pre-staging large numbers of operations) do not provide sustained, correct coordination under realistic NIC resource limits.

This is a significant shortcoming given the prevalence of such systems in modern HPC. For instance, among the ten highest-ranked systems in the November 2025 Top500~\cite{top500nov2025}, six deploy HPE Slingshot as their primary interconnect, including El Capitan, Frontier, and Aurora. Yet existing GPU communication frameworks do not provide a portable GPU-driven coordination model on these OFI-based fabrics. Vendor runtimes have begun exposing GPU-triggered interfaces, but these capabilities are limited in scope: they focus on point-to-point data movement and do not provide device-side collective coordination or kernel-level integration of distributed synchronization with GPU control flow. Consequently, GPUs on today's largest HPC systems still lack the ability to autonomously drive distributed coordination~\cite{sc25-gpuinit}.

This paper presents \textbf{GICC}, a GPU-driven distributed coordination framework designed explicitly for OFI-based interconnects such as HPE Slingshot, while also providing a streamlined path on GPU-native fabrics such as InfiniBand. Our primary systems contribution targets the OFI/CXI setting, where the coordination gap is most acute; on InfiniBand, GICC offers a lighter device-side path that exposes the same coordination interface but benefits from richer GPU--NIC integration. Even on GPU-native fabrics, existing GPU communication runtimes can incur avoidable device-side overheads on the critical path (e.g., shared-queue arbitration and redundant synchronization), motivating a coordination-oriented design that keeps device-side control lightweight. GICC exposes coordination as a first-class abstraction and enables GPU kernels to invoke distributed synchronization within kernel execution. GICC decouples coordination semantics from data movement mechanisms: GPUs drive coordination decisions and observe completion conditions, while the transport realizes the underlying data movement.

To remain correct and live under finite NIC resources, GICC employs a resource-aware execution model with a lightweight host monitor that performs asynchronous progress and resource reclamation when required by the interconnect. This keeps the host off the GPU fast path while ensuring that triggered NIC state can be safely recycled for repeated coordination.

We implement GICC on both NVIDIA and AMD GPUs and evaluate it over InfiniBand and Slingshot (OFI) interconnects. Our evaluation shows that GICC reduces per-coordination latency by up to $229\times$ compared to host-driven approaches and improves weak scaling efficiency by up to 25\%. On InfiniBand, GICC achieves up to $1.95\times$ lower put latency than NVSHMEM by eliminating unnecessary locking and synchronization. For an industrial stencil-based proxy application on 64 AMD MI250X GCDs, GPU-aware MPI incurs 52\% higher communication time than GICC. These results demonstrate that GPU-driven, resource-resilient distributed coordination constitutes a distinct and previously unaddressed design point in modern HPC systems.

In summary, this paper makes the following contributions:
\begin{itemize}[wide]
\item We characterize the structural limitations of existing GPU communication frameworks on OFI-based fabrics: NVSHMEM is host-mediated (not GPU-triggered), and no runtime provides portable GPU-driven coordination primitives on these dominant interconnects.
\item We design and implement a GPU-triggered communication layer over both GPU-native and host-managed interconnects, enabling GPU kernels to trigger NIC operations without host involvement on the fast path.
\item We introduce GPU-driven coordination primitives---including active messages and barriers---that enable GPU control flow to govern distributed synchronization. On OFI/CXI, we address finite NIC resource constraints via a hybrid progress model with asynchronous resource reclamation and a double-buffered stage-ahead mechanism driven by a lightweight host monitor.
\item We evaluate GICC on NVIDIA and AMD GPUs over both Slingshot and InfiniBand systems, demonstrating significant reductions in coordination overhead and communication time across microbenchmarks and real applications.
\end{itemize}

\section{Background}
\label{sec:background}

This section summarizes the execution and communication context of modern GPU-based distributed systems. We focus on (i) GPU execution and its interaction with synchronization boundaries, (ii) GPU-visible communication frameworks and their progress semantics, and (iii) the capabilities and resource limits exposed by high-performance NICs. These elements establish the technical context for the problem analysis in Section~\ref{sec:problem}.

\subsection{GPU Execution and Synchronization Boundaries}
\label{subsec:gpu_model}

GPUs execute programs as kernels comprising large numbers of lightweight threads. Modern GPU applications increasingly embed control flow within kernels, including phase-based execution, device-side conditionals, and iterative loops driven by intermediate results. Such patterns often require ordering between phases and synchronization across ranks to ensure correctness and enable overlap between computation and communication.

Current programming models separate kernel execution from distributed synchronization. GPU kernels execute without direct access to system-level collectives or global coordination services, and distributed coordination is typically initiated at kernel boundaries, once control has returned to the host. As a result, kernel boundaries serve as practical synchronization points between GPU computation and distributed execution.

\subsection{Interconnect Landscape and Progress Models}
\label{subsec:gpu_comm}

Several frameworks provide GPU-visible communication interfaces for distributed GPU applications. GPU-aware MPI allows communication operations to access GPU-resident buffers, and NVSHMEM provides a partitioned global address space abstraction that enables GPU threads to issue one-sided operations~\cite{nvshmem}. However, GPU visibility of an API does not by itself imply that ordering, completion, or collective synchronization can be driven from within GPU execution. These properties depend on the interconnect, its GPU--NIC integration model, and the progress semantics of the associated runtime.

\subsubsection{GPU-Native NIC Interfaces (e.g., InfiniBand)}

On platforms with GPU-native NIC interfaces, GPUs can directly trigger network work and observe completion with minimal host involvement. This enables designs in which GPU control flow can initiate, advance, and complete communication and coordination actions without relying on a host-driven progress engine on the fast path. As a result, GPU-initiated communication can more naturally compose with kernel-level control flow.

\subsubsection{OFI/libfabric Triggered Operations (e.g., Slingshot/CXI)}
\label{subsec:ofi_cxi}

On OFI-based systems such as HPE Slingshot, applications typically use libfabric~\cite{libfabric} with the CXI provider~\cite{libfabric-cxi-docs}. In this environment, in the general case, device code cannot dynamically submit new network work descriptors. Instead, communication and coordination must be expressed as host-prepared NIC work that can later be triggered by GPU activity.

Libfabric and the CXI provider expose a triggered execution model based on deferred work and NIC-side counters. Concretely, the host prepares a set of network operations and places them into a provider mechanism that defers execution (CXI's deferred work queue, or DWQ)~\cite{libfabric-cxi-docs}. Each deferred operation is associated with a trigger condition expressed in terms of a NIC counter and a threshold. When the counter reaches the programmed threshold, the NIC releases the corresponding deferred work and executes it.

From the GPU perspective, triggering is a lightweight doorbell write that updates a GPU-mapped NIC counter. Completion can be made GPU-visible by configuring the NIC to write to pre-registered GPU memory flags that GPU threads can poll. On CXI, the provider progress model is manual (\texttt{FI\_\allowbreak PROGRESS\_\allowbreak MANUAL})~\cite{libfabric-cxi-docs}, so applications typically drive libfabric progress from the host, e.g., by polling completion objects.

\paragraph{Terminology.}

\begin{figure}[t]
    \centering
    \includegraphics[width=.65\linewidth]{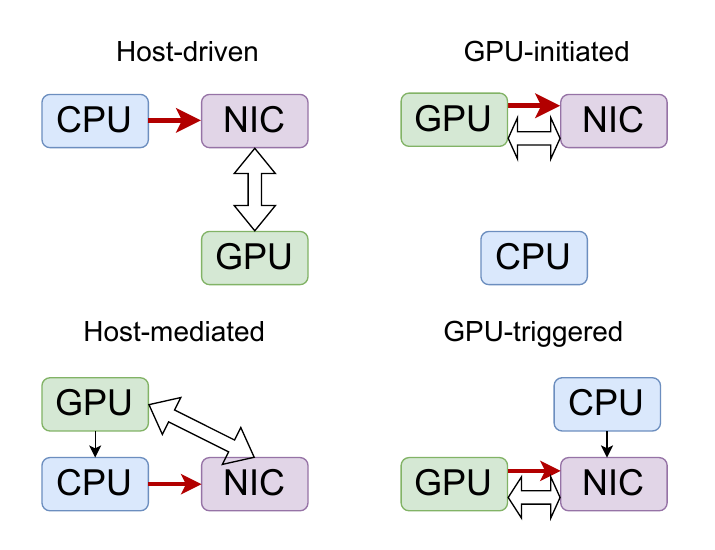}
    \caption{Four common communication modes between GPU and NIC. Red arrows indicate the triggering path, hollow double-headed arrows represent data movement, and black arrows denote other necessary operations.}
  \label{fig:3modes}
\end{figure}

Figure~\ref{fig:3modes} illustrates four common communication modes, where red arrows indicate the triggering path. Table~\ref{tab:capability} summarizes the communication modes adopted by various communication frameworks.
\textbf{Host-driven} means the host submits communication work and controls coordination decisions and progress.
\textbf{Host-mediated} means a GPU-visible API may exist, but the host still mediates progress and completion (e.g., host-managed queues/proxies), so GPU control flow cannot autonomously drive coordination.
\textbf{GPU-initiated} means GPU threads can invoke communication operations via a GPU-visible API; this does not imply GPU-driven progress or collective coordination semantics on all interconnects.
\textbf{GPU-triggered} means the host pre-stages a bounded set of NIC work and trigger conditions, and GPU kernels trigger execution (e.g., via doorbells/counter updates), so GPU control flow determines \emph{when} coordination actions occur on the fast path.

\begin{table}[t]
\centering
\caption{GPU coordination capabilities across runtimes and interconnects.}
\label{tab:capability}
\begin{tabular}{lcc}
\toprule
\textbf{Runtime} & \textbf{InfiniBand} & \textbf{Slingshot (OFI/CXI)} \\
\midrule
NVSHMEM~\cite{nvshmem}          & GPU-initiated       & Host-mediated \\
rocSHMEM~\cite{rocshmem}         & GPU-initiated       & Host-mediated \\
NCCL~\cite{nccl}      & GPU-initiated         & Host-driven \\
RCCL~\cite{rccl}      & Host-driven         & Host-driven \\
GPU-aware MPI\footnotemark~\cite{mpi50} & Host-driven & Host-driven \\
GICC (this work) &  GPU-initiated         & GPU-triggered \\
\bottomrule
\end{tabular}
\end{table}

\footnotetext{We use \emph{GPU-aware MPI} to denote MPI implementations that accept GPU-resident buffers as arguments to standard (host-side) MPI calls; the MPI API and semantics are defined by the MPI 5.0 standard~\cite{mpi50}. The MPI 5.0 standard does not define any GPU-initiated/device-side MPI operations; vendor-specific extensions that expose deeper GPU/NIC capabilities are discussed in Section~\ref{sec:implementation}.}

\subsection{NIC Resource Model}

Triggered NIC mechanisms provide low-latency execution of pre-staged operations, but they consume finite NIC-resident resources. In particular, provider-managed structures such as deferred work queues, NIC counters, and completion tracking state bound how many operations can be armed concurrently and how frequently coordination can be repeated. Repeated GPU-triggered collectives must therefore implement a safe lifecycle for NIC state reuse (retire $\rightarrow$ reset $\rightarrow$ re-arm) and avoid assuming unbounded outstanding work. In Section~\ref{sec:problem}, we quantify how these limits manifest on CXI and show that naive pre-staging quickly leads to resource exhaustion and induces blocking.

\section{Problem Analysis}
\label{sec:problem}

This section shows why host-driven coordination becomes a dominant bottleneck
in modern GPU-based distributed workloads and why existing GPU-visible
communication mechanisms are insufficient to express \emph{GPU-driven} distributed
coordination on OFI-based interconnects. Our focus is not on the efficiency of
any particular implementation, but on structural limitations arising from kernel
boundary transitions, progress semantics, and finite NIC resources.

\subsection{Quantifying the Host-Driven Coordination Bottleneck}
\label{subsec:quantify}

We consider a minimal phase-based workload that repeatedly alternates between a
GPU compute segment and a global coordination step. The \emph{total} compute work
is held constant, while the coordination frequency is increased by partitioning
the computation into $N$ phases. This pattern captures common structures in
iterative solvers, pipelined stages, and multi-phase GPU applications, where
global ordering or synchronization occurs frequently.

\begin{figure}[t]
  \centering
  \includegraphics[width=0.85\linewidth]{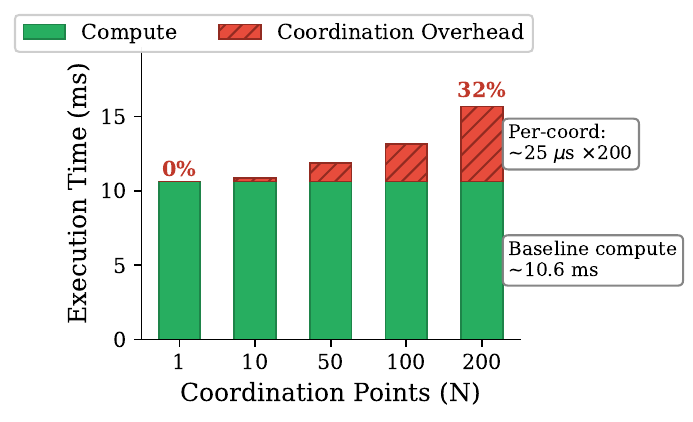}
  \caption{Execution time breakdown of a phase-based GPU workload with kernel-boundary, host-driven coordination, for increasing numbers of coordination phases $N$.}
  \label{fig:bottleneck}
\end{figure}

Figure~\ref{fig:bottleneck} reports a time breakdown for a host-driven
implementation. We measure the execution time of a fixed workload
($10^{11}$ floating-point operations) distributed across $N$ coordination
phases on two nodes equipped with AMD MI250X GPUs and a Slingshot
interconnect. Each phase consists of three steps: (1) launching a HIP kernel
that performs floating-point computation, (2) calling
\texttt{hipDeviceSynchronize()} to wait for kernel completion, and (3)
executing \texttt{MPI\_Barrier()} to enforce inter-node synchronization.
The total amount of computation is held constant by dividing the work evenly
across phases.

As $N$ increases, end-to-end execution time rises substantially even though
the baseline compute time remains approximately constant (about 10.6~ms).
The stacked bars show that the increase in total execution time is primarily
attributable to coordination overhead introduced at each kernel boundary,
rather than additional GPU computation. The figure further indicates a
roughly constant per-coordination cost (on the order of tens of microseconds),
which accumulates linearly with $N$ and constitutes an increasingly large
fraction of total runtime as coordination becomes more frequent.

These results indicate that frequent coordination is bottlenecked primarily by
kernel-boundary round trips and host-driven progress, not by the payload size of
the coordination itself. This motivates the need for mechanisms that allow GPUs
to participate in distributed coordination \emph{within} GPU execution, without
re-entering host control flow.

\subsection{OFI Progress Gap Under Manual Progress}
\label{subsec:ofi_progress_gap}

On OFI-based systems such as Slingshot (CXI), the GPU can trigger \emph{pre-staged} work and observe NIC writebacks into GPU memory, but it cannot enqueue new work descriptors dynamically (Section~\ref{subsec:ofi_cxi}). More subtly, GPU-visible completion is not equivalent to provider-visible progress and resource retirement under libfabric manual progress. In practice, repeated GPU-triggered coordination requires host-side progress and explicit re-arming of provider-managed state; otherwise, applications eventually stall as resources are exhausted.

\begin{figure*}[t]
  \centering
  \includegraphics[width=\linewidth]{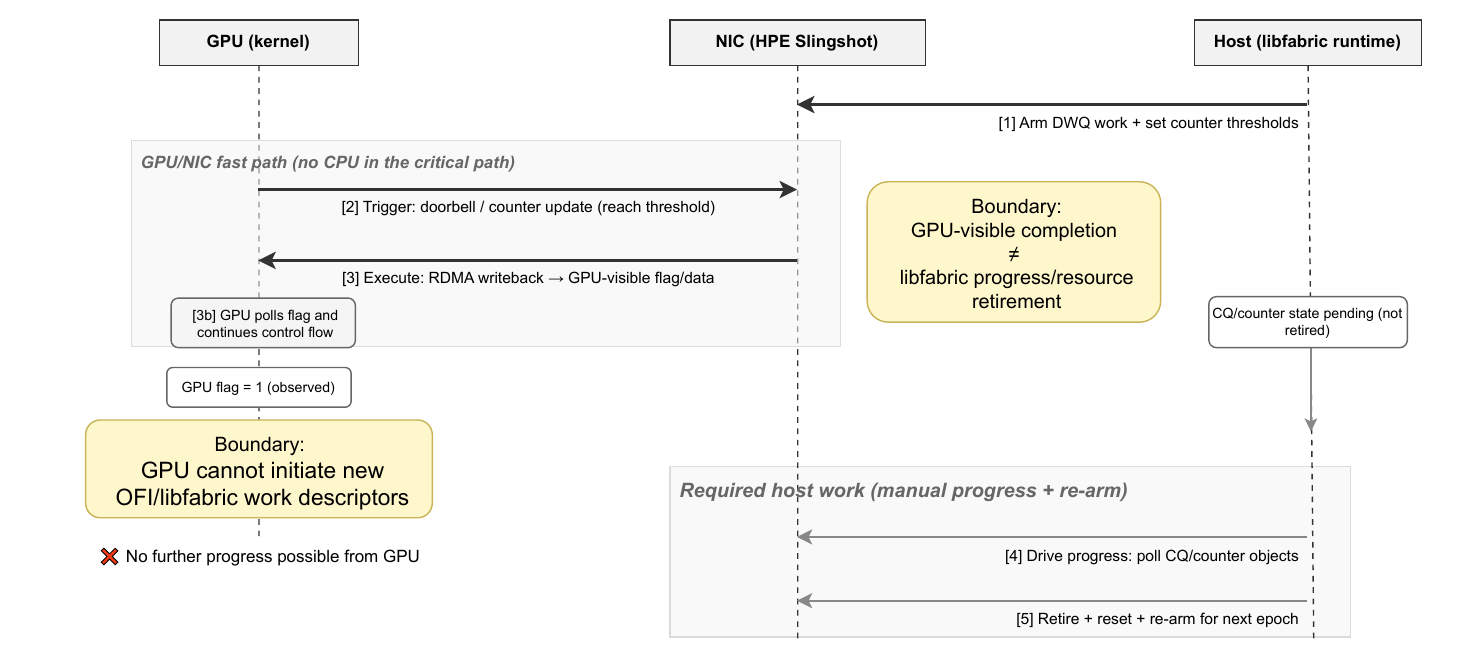}
  \caption{Boundary conditions for GPU-driven coordination on Slingshot (CXI) with libfabric manual progress. The host first arms deferred NIC work (DWQ) and programs counter thresholds (1). A GPU kernel can then trigger execution via a doorbell/counter update (2), after which the NIC executes the deferred operations and writes completion signals into GPU-visible memory (3), enabling the GPU to poll flags and proceed within kernel control flow. However, GPU-visible completion is not equivalent to libfabric progress or resource retirement: under \texttt{FI\_PROGRESS\_MANUAL}, the host must drive progress by polling completion objects (4) and then retire/reset and re-arm resources for the next epoch (5).}
  \Description{Sequence diagram with three lifelines (GPU kernel, NIC CXI, and host libfabric runtime). Steps show: host arms deferred work and counter thresholds; GPU triggers via doorbell/counter update; NIC performs RDMA writeback to GPU-visible flags; GPU polls flags; host drives manual progress by polling CQ/counter objects and then retires, resets, and re-arms for the next epoch. Callouts highlight that GPUs cannot initiate new OFI work descriptors and that GPU-visible completion does not imply libfabric progress/resource retirement.}
  \label{fig:cxi_boundary}
\end{figure*}




\subsection{Resource Exhaustion Under Finite NIC State}
\label{subsec:nic_state}

GPU-triggered coordination relies on triggered work, counters, and completion tracking maintained by the NIC and provider. This state is finite, and under frequent coordination it can become a first-order constraint. To make these limits concrete, we examine the CXI provider on HPE Slingshot through documentation review and experimental measurement.

\begin{table}[t]
  \centering
  \small
  \begin{tabular}{rccc}
  \toprule
  Ranks ($P$) & DWQ/barrier ($R$) & Counter inc./barrier ($2R$) & Max pre-staged \\
  \midrule
  64   & 6  & 12 & 42 \\
  256  & 8  & 16 & 32 \\
  1024 & 10 & 20 & 25 \\
  4096 & 12 & 24 & 21 \\
  \bottomrule
  \end{tabular}
  \caption{Maximum barrier instances that can be pre-staged on CXI, 
  constrained by DWQ capacity (256 entries) and counter range (2047), where the maximum is $\min(\lfloor 256/R\rfloor, \lfloor 2047/(2R)\rfloor)$ for $R=\lceil \log_2 P \rceil$.}
  \label{tab:resource}
\end{table}

\paragraph{NIC Resource Limits.}
The HPE Slingshot 11 NIC exposes several key resource constraints for triggered operations.
First, each trigger counter is bounded to a maximum value of 2047, and this bound is shared across all ranks that use the same NIC.
Second, the deferred work queue (DWQ) can hold at most 256 entries before requiring a flush.
Finally, a single triggered communication consumes two counters (e.g., a trigger counter and a completion/accounting counter). These limits are hardware constraints that cannot be removed through software configuration.

\paragraph{Resource Consumption Analysis.}
For a dissemination barrier across $P$ ranks, each barrier instance requires 
$R = \lceil \log_2 P \rceil$ communication rounds, with each round consuming 
two counter increments and one DWQ entry. Table~\ref{tab:resource} summarizes 
the resulting per-barrier resource consumption at representative scales.

At $P = 64$, a single barrier requires 6 DWQ entries and 12 counter increments, so the host can pre-stage at most 42 barrier instances before DWQ capacity is exhausted. At larger scales ($P = 4096$), this drops to 21 instances. Since the counter limit is shared across ranks on the same NIC, the effective pre-staging budget per rank decreases as the number of ranks sharing a NIC grows.

\paragraph{The Blocking Flush Problem.}
When DWQ is full or counters reach their maximum value, the CXI provider requires a flush operation to reset counters and retire DWQ entries. Critically, the current CXI implementation implements flush conservatively: it includes a blocking wait on the order of one second to ensure that in-flight operations have completed. Such behavior is entirely unsuitable for high-frequency coordination. An alternative approach is to poll completion queues from a host thread via \texttt{fi\_cq\_read} to update resource state, but this incurs significant overhead and often becomes a bottleneck over repeated iterations.

These constraints motivate a key requirement for GPU-driven coordination: mechanisms must bound NIC-resident state and provide forward progress without accumulating unbounded triggered work. This limitation is fundamental---realistic NICs cannot represent arbitrary numbers of outstanding coordination events directly in hardware.

\subsection{Why Existing Approaches Fall Short}
\label{subsec:existing_fail}

Persistent kernels and polling are common techniques for reducing kernel launch overhead by keeping GPU threads resident and spinning on GPU-visible flags. While this can eliminate kernel-boundary transitions, it does not resolve the coordination problem on OFI fabrics. First, persistent polling consumes GPU execution resources and competes with computation, reducing effective overlap. Second, GPUs cannot enqueue new OFI work descriptors; device code can only trigger work that has been prepared in advance by the host (Section~\ref{subsec:ofi_cxi}). Supporting repeated coordination would therefore require either pre-staging an unbounded number of operations---impossible under finite NIC resources---or host-side re-arming between coordination instances.

Moreover, existing GPU-visible communication frameworks such as NVSHMEM expose data movement primitives (put/get); however, on OFI fabrics, completion and ordering remain host-managed. Consequently, GPU threads can initiate transfers but cannot autonomously determine when coordination conditions are satisfied or when it is safe to proceed to the next phase.

\section{Design of GICC}
\label{sec:design}

GICC is designed to enable GPU-driven distributed coordination across heterogeneous interconnect environments. While our primary motivation is to close the coordination gap on OFI-based fabrics such as HPE Slingshot, the design of GICC is not specific to OFI. We implement and evaluate GICC on both InfiniBand systems with GPU-native NIC interfaces and OFI-based systems with host-managed NICs. This section presents a unified design view that highlights which aspects are fundamental to GPU-driven coordination and which arise specifically from OFI constraints.

\begin{figure}[t]
\begin{lstlisting}[
  language=C++,
  basicstyle=\ttfamily\scriptsize,
  keywordstyle=\color{blue}\bfseries,
  commentstyle=\color{gray}\itshape,
  stringstyle=\color{red},
  numbers=left,
  numberstyle=\tiny\color{gray},
  frame=single,
  breaklines=true,
  morekeywords={__global__,__device__,__syncthreads}
]
// ===== Host: pre-stage two puts with increasing thresholds =====
void host_prestage(gicc_trigger_t trig, int left, int right) {
  // Put to left neighbor, fires when counter >= 1
  gicc_prestage_put(trig, threshold=1,
      dest=left_halo, src=left_bnd, size=N, pe=left);
  // Put to right neighbor, fires when counter >= 2
  gicc_prestage_put(trig, threshold=2,
      dest=right_halo, src=right_bnd, size=N, pe=right);
}

// ===== GPU: single counter update releases both puts =====
__global__ void halo_kernel(float *data, uint64_t *sig,
                            int ny, gicc_trigger_t trig) {
  int tid = threadIdx.x + blockIdx.x * blockDim.x;
  
  if (tid < ny) data[tid] = compute(tid);
  __syncthreads();
  
  if (tid == 0) {
    // Set counter to 2: both threshold-1 and threshold-2 fire
    gicc_trigger(trig, counter_val=2);
    
    // Poll GPU-visible flags (GE = greater or equal)
    // You could also use gicc_quiet() here
    gicc_wait_until(&sig[0], GICC_CMP_GE, 1);  // left done
    gicc_wait_until(&sig[1], GICC_CMP_GE, 1);  // right done
  }
  __syncthreads();
  
  if (tid < ny) stencil_update(data, tid);
  if (tid == 0) gicc_barrier_all();
}
\end{lstlisting}
\caption{GICC \textit{put} example (pseudocode) on OFI/CXI: the host pre-stages two \textit{put} operations into the deferred work queue (DWQ) with thresholds 1 and 2 on the same trigger counter; a single GPU counter update to 2 releases both simultaneously. \texttt{gicc\_wait\_until} polls GPU-visible flags using comparison operators (e.g., \texttt{GICC\_CMP\_GE} for $\geq$).}
\label{fig:gicc-ofi}
\end{figure}

\subsection{Design Objectives}

GICC is guided by three objectives. First, coordination decisions should be expressible and triggerable from within GPU execution. GPU kernels should be able to decide when a synchronization point or coordination event occurs and initiate distributed coordination without terminating kernels or returning control to the host on the fast path.

Second, coordination and communication should incur minimal latency and CPU involvement. GPU-driven coordination must enable tight compute--communication overlap and avoid unnecessary synchronization, locking, or control-path overheads.

Third, the design must remain correct and viable under finite NIC resources. On fabrics where NIC state is host-managed or limited, coordination mechanisms must avoid unbounded accumulation of outstanding work and provide forward progress.

\subsection{Coordination Interface and Scope}

GICC exposes GPU-visible coordination primitives that can be invoked from within a kernel. Conceptually, the coordination interface separates (i) \emph{triggering} a coordination action and (ii) \emph{accepting and synchronizing} on triggered actions. This decomposition addresses the shortcomings identified in Section~\ref{sec:problem}: GPUs should be able to initiate coordination without host round trips, while progress and state must remain bounded under finite NIC resources (Section~\ref{subsec:nic_state}).

From an application developer's perspective, this split also keeps the integration model compact. The host is responsible for creating communication endpoints, registering GPU-visible buffers and signal locations, and, on OFI/CXI, pre-staging the deferred NIC work associated with a coordination instance. Inside the kernel, a small number of designated control threads trigger the prepared action, poll GPU-visible completion state, and then release the rest of the kernel to proceed. In Figure~\ref{fig:gicc-ofi}, the data buffers to be communicated are specified during the host-side pre-staging calls (\texttt{src}, \texttt{dest}, \texttt{size}, and peer identifier), while the GPU-side call determines \emph{when} those already-armed transfers should execute.

\textbf{Trigger.} Triggering initiates a coordination action from GPU control flow and is designed to be lightweight and asynchronous (i.e., it does not require GPU-side blocking beyond local bookkeeping). In GICC, triggering is realized through a GPU-triggered \textbf{active message} path; we describe the execution model and how triggers release NIC work in the next subsection.

\textbf{Accept and synchronize.}
Acceptance and synchronization complete coordination actions by (a) receiving and dispatching coordination messages and (b) enforcing ordering and completion conditions. Both GPU and CPU may participate. On the receive side, GPU kernels can poll GPU-visible mailboxes or flags for the arrival of an active message and then dispatch the corresponding handler within kernel control flow. Symmetrically, a host-side progress thread can poll and accept the same triggered events, providing a portable fallback path and assisting with progress and resource retirement on fabrics with manual progress.
To support iterative GPU workloads, synchronization and collective operations are designed to be lightweight, asynchronous, and resource-aware. GICC does not introduce warp- or block-level collectives. Intra-kernel synchronization is handled using existing CUDA/HIP primitives such as \texttt{\_\_syncthreads()}. The interface does not restrict which GPU threads may invoke coordination primitives; however, applications typically designate a small number of control threads to avoid redundant triggering. This is an application-level convention rather than an API requirement.





\subsection{GPU-Triggered Coordination Model}

At the core of GICC is a GPU-triggered execution model. Coordination is decomposed into two roles: preparation of network work descriptors and triggering of their execution. On OFI/CXI, GPU kernels do not enqueue network operations directly; instead, work is prepared in advance by the runtime and executed when triggered by GPU activity. This structure allows GPU control flow to determine \emph{when} coordination occurs, while keeping network execution on the fast path independent of host control flow. On InfiniBand systems with GPU-native NIC access, the same coordination interface maps to direct GPU posting rather than host-prepared deferred work.

An important property of this model is that a single GPU trigger may release a sequence of pre-staged network operations. This aggregation amortizes GPU--NIC interaction overhead across multi-step coordination sequences (e.g., multi-round barriers) and across dense communication patterns (e.g., fan-out notifications), reducing GPU triggering overhead to a constant number of doorbell actions per coordination instance. 

Figure~\ref{fig:gicc-ofi} illustrates this model concretely on OFI/CXI, where NIC work must be host-managed: the host pre-stages two \texttt{put} operations with increasing thresholds on a shared trigger counter; at runtime, a single GPU counter update releases both operations, and the kernel polls GPU-visible flags for completion before proceeding. On InfiniBand systems with GPU-native NIC access, the pre-staging phase is unnecessary---GPU kernels can directly construct and submit RDMA operations via NIC doorbells without host involvement. Section~\ref{sec:implementation} details both realizations.

\subsection{Hybrid Progress and Handoff}

On fabrics where GPUs cannot fully manage NIC state (e.g., OFI-based systems), GICC employs a hybrid progress model that separates coordination authority from liveness and resource management.

GPU threads trigger coordination and observe completion via GPU-visible memory updates. A lightweight host monitor thread operates outside the GPU fast path to (i) drive provider-visible progress when required and (ii) perform resource retirement and re-arming needed for repeated execution. Importantly, the host does not participate in coordination decisions or define ordering; it only enforces liveness and safe reuse of bounded NIC state.

Handoff between GPU execution and host re-arming is explicit: GPU execution proceeds only when the runtime has ensured that the next instance is fully armed and safe to trigger. This prevents coordination logic from assuming unbounded outstanding NIC state while keeping the host off the fast path. The concrete handoff mechanism is described in Section~\ref{sec:implementation}.

On CXI in particular, libfabric progress is manual (\texttt{FI\_\allowbreak PROGRESS\_\allowbreak MANUAL})~\cite{libfabric-cxi-docs}, so host-side polling of completion objects is required to drive provider-visible progress and resource retirement. Thus, even when coordination authority resides in GPU control flow, GICC retains a minimal host monitor to ensure liveness and to manage the lifecycle of triggered work.

\subsection{Coordination versus Communication}

GICC separates coordination semantics from communication mechanisms. Coordination defines ordering and completion conditions, while communication primitives (e.g., RDMA writes and doorbells) are used to realize these semantics efficiently. This separation allows coordination logic to be shared across interconnects while enabling transport-specific optimizations on platforms that expose richer GPU--NIC interfaces.



\section{Implementation}
\label{sec:implementation}

We implement GICC on both InfiniBand systems with GPU-native NIC interfaces and OFI-based systems targeting HPE Slingshot (CXI).

\subsection{InfiniBand Implementation}

On InfiniBand systems, GPUs directly manage network operations using GPU-accessible NIC doorbells and queues. This allows GPUs to construct work queue elements, trigger doorbells, and poll completion queues without host intervention. As a result, many challenges that arise on OFI-based systems---such as host-managed NIC state, triggered work re-arming, and resource reclamation---do not apply.

On NVIDIA/Mellanox \texttt{mlx5} systems, this GPU-native path is enabled by exposing NIC MMIO doorbells to the GPU via the User Access Region (UAR). GPU threads can directly construct InfiniBand Work Queue Elements (WQEs) in device memory and ring the BlueFlame doorbell to trigger the NIC DMA engine, yielding a fully device-side submission path for RDMA operations.

We implement active messages and coordination primitives directly on top of GPU-managed RDMA operations. Active messages on InfiniBand use the same receive-side semantics as in the OFI implementation: NICs deliver message payloads via RDMA writes into GPU memory, and GPU kernels poll for arrival and dispatch handlers. To keep the fast path lightweight, our InfiniBand backend uses an exclusive-QP (single-owner) posting model to eliminate shared-QP arbitration, and it polls the CQE owner bit for completion detection. We further leverage InfiniBand's in-order completion semantics within a QP to implement active messages via a simple sequence protocol: a sender writes the payload before a final sequence update, and the receiver polls the sequence field to detect message readiness.

\subsection{OFI/CXI Implementation}
\label{subsec:ofi_impl}

On OFI-based systems such as Slingshot, NIC work descriptors must be enqueued by the host using \texttt{fi\_control(FI\_QUEUE\_WORK)}. GICC prepares triggered RDMA operations on the host and exposes NIC trigger counters to the GPU via vendor-specific extensions. GPU kernels trigger execution by writing threshold values to these counters via MMIO.

\paragraph{Triggered execution.}
GICC represents GPU-triggerable actions as host-queued deferred work (DWQ entries) guarded by NIC counter thresholds (Section~\ref{subsec:ofi_cxi}). The host associates each deferred operation with a threshold on a GPU-visible trigger counter. A GPU kernel then advances the counter (via a doorbell write) to release the corresponding operations. By arming a sequence of operations with increasing thresholds, a single device-side counter update can release multiple pre-staged operations, amortizing GPU triggering overhead across multi-step coordination sequences.

\paragraph{Host monitor and epoch-based handoff.}
Because CXI uses libfabric manual progress (\texttt{FI\_\allowbreak PROGRESS\_\allowbreak MANUAL})~\cite{libfabric-cxi-docs}, GICC maintains a lightweight host monitor thread to drive provider-visible progress and to retire and re-arm triggered state. Handoff between GPU execution and host re-arming is explicit: each coordination instance is assigned a monotonically increasing epoch, and GPU threads trigger an instance only after observing that a GPU-visible readiness counter has reached the expected epoch. The host increments this counter only after completing provider progress and fully arming the next instance. This mechanism guarantees safe reuse of bounded NIC state without placing the host on the GPU fast path. In our measurements, the CPU thread introduces an average overhead of approximately $1\,\mu\text{s}$ per communication operation.

\paragraph{Double-buffered DWQ stage-ahead for barrier primitives.}
On OFI/CXI, GPU-triggered barriers introduce an inherent CPU--GPU dependency: the host must enqueue DWQ descriptors before the GPU can trigger them. If barriers occur more frequently than the host can prepare descriptors, GPUs stall waiting for DWQ setup.
To decouple host-side preparation from GPU-side execution, GICC double-buffers DWQ descriptor preparation. GICC maintains two rotating signal slots and source buffers per round, with barrier $i$ mapping to slot $i \bmod 2$: while the GPU triggers and polls slot $(i{-}1) \bmod 2$ for barrier $i{-}1$, the host monitor has already retired slot $i \bmod 2$ and re-armed it for barrier $i$.
The two-slot structure serves as a transition buffer between consecutive barriers, not as a window of concurrent in-flight barriers: dissemination rounds remain strictly sequential and only one barrier is actively triggered at a time. This stage-ahead pattern hides per-barrier DWQ setup latency behind GPU barrier execution and yields constant host work per barrier. At any instant the DWQ holds at most $2R$ armed operations per rank, where $R = \lceil \log_2 P \rceil$ is the number of dissemination rounds, so pre-staged state never grows with the application's barrier count. Safe reuse between generations is enforced by the epoch-based handoff described above.
Because the stage-ahead pattern is oblivious to the specific collective it drives, the monitor+double-buffer structure generalizes naturally to other triggered collectives built on top of GICC.

\paragraph{Avoiding blocking flush.}
Section~\ref{subsec:nic_state} shows that CXI exposes tight bounds on counter ranges and DWQ capacity, and that provider-level flush can be extremely expensive. By construction, GICC avoids reaching these failure modes in the steady state. First, the total pre-staged barrier work is bounded: at any instant the host monitor keeps at most $2R$ DWQ entries armed per rank, where $R=\lceil \log_2 P \rceil$ is the dissemination round count. Even at extreme scales (e.g., $P=10^9$ giving $R=30$), this fits within CXI's DWQ and counter budgets regardless of how many barriers the application invokes. Second, the host monitor continuously retires completed work and re-arms window slots under epoch handoff, preventing unbounded accumulation of armed state. If resource pressure still arises (e.g., due to unexpected delays in provider progress), GICC falls back to a host-issued, non-triggered path (Section~\ref{subsec:defensive_fallback}) to preserve correctness while relieving pressure.

\subsection{Barriers and Active Messages}

The barrier primitive is implemented as a multi-round dissemination protocol on both fabrics. For $P$ ranks, a barrier consists of $R=\lceil \log_2 P \rceil$ rounds; in each round, the NIC performs an RDMA write to a GPU-visible signal location on a peer rank. GPU threads poll these locations to detect round completion before advancing. On OFI, these RDMA actions are executed as host-queued DWQ descriptors and released by GPU-visible trigger counters; the double-buffered stage-ahead described in Section~\ref{subsec:ofi_impl} keeps the host monitor one barrier ahead of the GPU to avoid CPU--GPU setup stalls. On InfiniBand, the same round operations are posted and triggered entirely by the GPU via doorbells.

Active messages are implemented uniformly across both InfiniBand and OFI fabrics, 
sharing a common message format and synchronization protocol while leveraging the 
platform-specific triggering mechanisms described above. Both implementations adopt 
a compact AM slot structure containing a sequence number, a message header 
(handler ID, source rank, and flags), and a user argument region. The sequence 
number is placed at the beginning of the slot and serves as an atomic release 
indicator. Message transmission is accomplished via two RDMA writes: the sender 
first writes the message body (header and arguments), then writes the sequence 
number as a release point. Due to RDMA in-order completion semantics within a
connection, the receiver only needs to poll the sequence field---when the observed
value matches the expected sequence number, the message body is guaranteed to
have arrived completely, eliminating the need for separate completion notifications
or additional synchronization overhead.

On the receive side, GPU kernels poll pre-registered mailbox slots for message 
arrival by checking sequence numbers, and dispatch handlers within kernel control 
flow upon detection. Both implementations share three key design principles that 
enable low-latency GPU-native messaging: (1) a fully GPU-autonomous communication 
path that eliminates CPU intervention on the fast path, (2) a lightweight 
synchronization mechanism based on RDMA ordering semantics rather than explicit 
completion events, and (3) a lock-free, single-owner resource model that avoids 
arbitration overhead. This unified abstraction provides a portable programming 
interface for GPU-native active messages across heterogeneous HPC interconnects.




\subsection{Defensive Fallback and Scope}
\label{subsec:defensive_fallback}

On OFI-based systems, GICC includes a defensive fallback path for unexpected resource pressure. The fallback is triggered when the runtime cannot safely enqueue or re-arm the triggered work for the next instance---for example, because provider-visible progress lags or a window slot cannot yet be retired. In that case, the host temporarily issues the corresponding non-triggered RDMA operations while GPU threads continue to wait on the same GPU-visible completion flags. Ordering semantics are preserved because the host issues the same communication sequence for that instance and only resumes GPU-triggered execution after the affected slot has been fully retired and re-armed. This path is intended as a correctness-preserving fallback and is not performance-critical.

The current implementation assumes fail-stop execution and reliable RDMA transport. Handling node failures, message loss, and dynamic membership is outside the scope of this work.

\section{Evaluation}
\label{sec:evaluation}

This section evaluates GICC across two representative GPU cluster platforms and compares it against state-of-the-art host-driven and GPU-aware communication runtimes. Our evaluation focuses on three aspects: (1) point-to-point communication latency, (2) active-message-based fine-grained communication efficiency, and (3) coordination overhead in iterative GPU workloads. We further demonstrate the end-to-end impact of GPU-driven coordination on matrix multiplication, a Jacobi stencil, and an industrial proxy application.

\subsection{Experimental Platforms}

We conduct experiments on two distinct platforms that represent different interconnect and NIC integration models. Tioga, operated by Livermore Computing, is equipped with four AMD MI250X GPUs (eight GCDs, i.e., eight logical GPUs) and four HPE Cassini NICs, connected via HPE Slingshot~11 interconnect. The software stack includes LibFabric~2.1, Cray MPICH~9.0.1, and GASNet~2025.8.0. Although Cray MPICH advertises kernel-triggered capabilities, we were unable to enable this mode on any of our tested Slingshot-based systems, including ALCF Polaris, NERSC Perlmutter, and Tioga. Consequently, we use GPU-aware Cray MPICH as the practical host-driven baseline in our evaluation. Maple, operated by TotalEnergies, features an NVIDIA GH200 GPU with a Mellanox ConnectX-7 NIC over InfiniBand HDR, running MLNX\_OFED~24.07, NVSHMEM~v3.3.24, and GASNet~2025.8.0. Experiments that illustrate the shortcomings described in Section~\ref{sec:problem} were conducted on Tioga. All experiments are conducted at least five times, and the figures report the mean with standard deviation. In our discussion below, we focus on central trends; the observed variance was not large enough to change the ranking of methods in the reported cases. On Maple, we configure NVSHMEM in IBGDA mode.

\subsection{Microbenchmarks}

We evaluate GICC using a suite of microbenchmarks on Tioga and Maple, covering coordination latency, point-to-point latency, and active message ping-pong performance.

\begin{figure}[t]
    \centering
    \includegraphics[width=\linewidth]{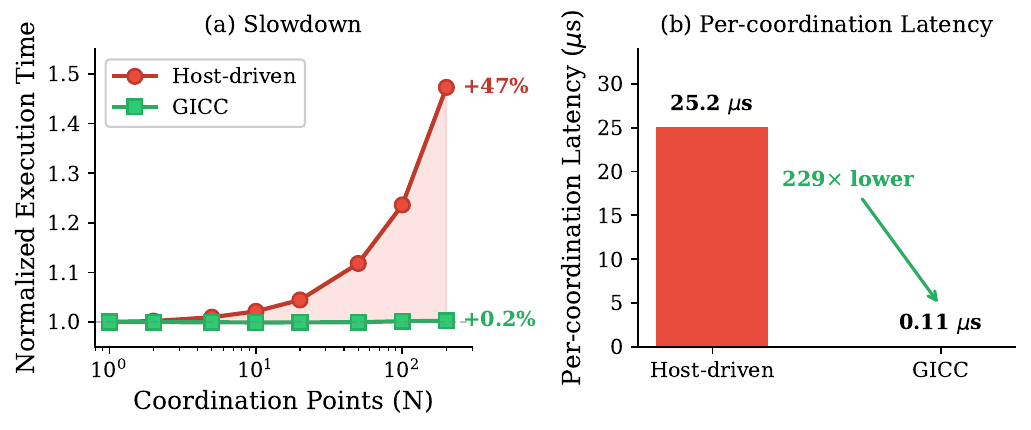}
    \caption{Coordination microbenchmark with fixed computation and varying numbers of coordination points $N$ on Tioga (HPE Slingshot + AMD MI250X). (a) End-to-end slowdown, normalized to the smallest coordination count. (b) Average per-coordination latency.}
  \label{fig:evaluation}
\end{figure}

Figure~\ref{fig:evaluation} isolates coordination overhead by fixing the total computation while varying the number of coordination points $N$. This microbenchmark uses the same workload and computation setup as in Section~\ref{subsec:quantify}: a fixed $10^{11}$-FLOP workload executed on two nodes. In Figure~\ref{fig:evaluation}(a), slowdown denotes end-to-end runtime normalized to the baseline runtime at the smallest coordination count. The host-driven baseline exhibits increasing end-to-end slowdown as $N$ grows, because each coordination point occurs at a kernel boundary and requires a CPU/runtime round-trip for synchronization and manual progress. In contrast, GICC keeps execution time nearly flat by enabling GPU-driven coordination with device-side progress. Figure~\ref{fig:evaluation}(b) reports a $229\times$ reduction in average per-coordination latency (from $25.2\,\mu\mathrm{s}$ to $0.11\,\mu\mathrm{s}$), explaining the widening gap in Figure~\ref{fig:evaluation}(a).

\begin{figure}[t]
    \centering
    \begin{minipage}[t]{0.49\linewidth}
        \centering
        \includegraphics[width=\linewidth]{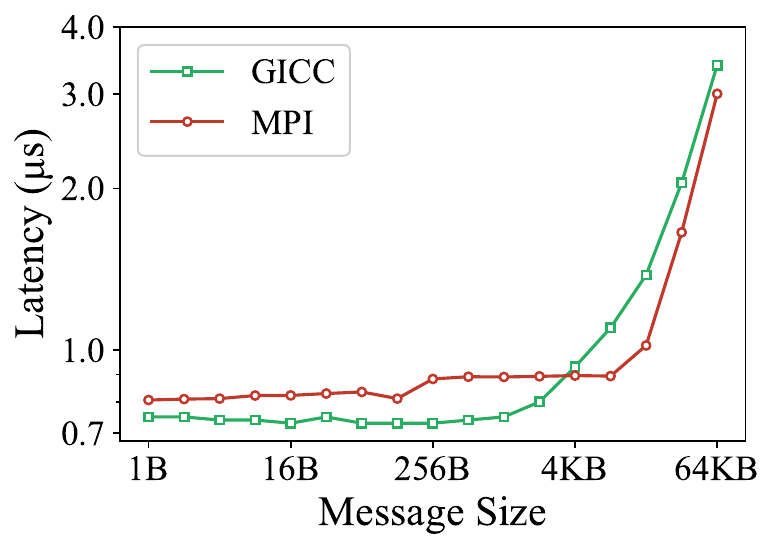}
        \captionof{figure}{P2P \texttt{put} latency on Tioga (HPE Slingshot 11 + AMD MI250X).}
        \label{fig:put_bw_A}
    \end{minipage}
    \hfill
    \begin{minipage}[t]{0.49\linewidth}
        \centering
        \includegraphics[width=\linewidth]{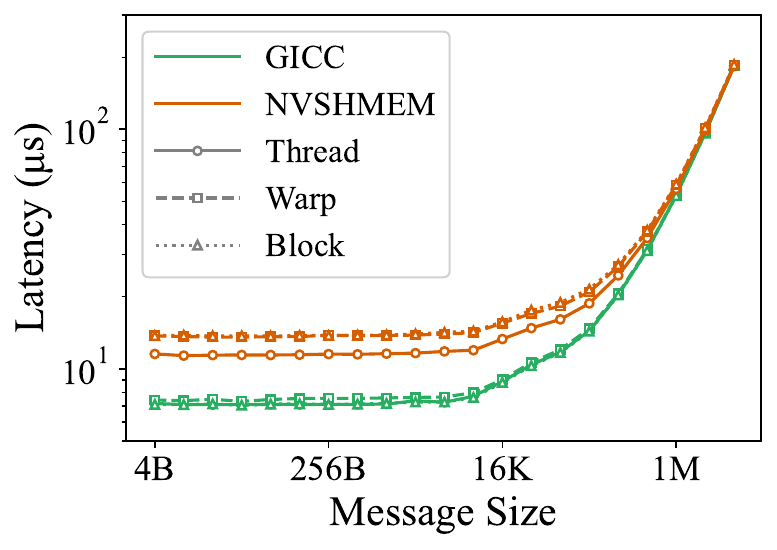}
        \captionof{figure}{P2P \texttt{put} latency across different scopes on Maple (InfiniBand HDR + GH200).}
        \label{fig:put_latency_B}
    \end{minipage}
\end{figure}

Figures~\ref{fig:put_bw_A} and~\ref{fig:put_latency_B} present concurrent \textit{put} latency microbenchmarks on Tioga and Maple, respectively. Here, latency refers to the elapsed time to complete one end-to-end \textit{put} operation under the synchronization protocol of the corresponding benchmark. Similar experiments were conducted for \textit{get} operations and yielded consistent trends; further detailed results are omitted for brevity.

On Tioga (Figure~\ref{fig:put_bw_A}), GICC is compared against Cray MPICH using the OSU Micro-Benchmarks (OMB) one-sided test. The benchmark launches 32 concurrent HIP streams, each issuing \textit{put} operations to a remote rank. For small messages ($\leq$2~KB), GICC achieves 7--14\% lower latency due to reduced host-side synchronization overhead. As message sizes grow, the transfer cost becomes dominated by network transmission time, causing both implementations to converge.

On Maple (Figure~\ref{fig:put_latency_B}), GICC is evaluated against NVSHMEM using the official \texttt{perftest} suite, specifically \texttt{put\_pingpong\_latency}, which measures the round-trip latency of repeated \texttt{shmemput} operations between two PEs under a ping-pong synchronization pattern. For small messages (4~B--1~KB), GICC achieves a latency of approximately $7.1\,\mu\mathrm{s}$, delivering a $1.62\times$--$1.94\times$ reduction (i.e., $38$--$49\%$ lower) compared to NVSHMEM's $11.5$--$13.8\,\mu\mathrm{s}$, depending on the synchronization scope. This improvement stems from GICC's streamlined WQE construction path that bypasses NVSHMEM's internal proxy thread and software queueing mechanisms. As message sizes increase beyond 8~KB, the performance gap narrows as network bandwidth becomes the dominant factor; at 4~MB, both implementations converge to approximately $185\,\mu\mathrm{s}$, achieving near line-rate throughput of 21--22~GB/s. Notably, GICC exhibits consistent latency across thread, warp, and block scopes (within $0.5\,\mu\mathrm{s}$), whereas NVSHMEM incurs an additional $2\,\mu\mathrm{s}$ overhead for collective scopes due to its internal synchronization barriers.

\begin{table}[H]
  \centering
  \caption{Short AM microbenchmark comparison between GICC and GASNet-EX on Tioga (HPE Slingshot 11 + AMD MI250X) and Maple (InfiniBand HDR + GH200), under two modes: ReqReq and ReqRep. Reported values are latencies in $\mu$s (lower is better).}
  \label{tab:short_am_gasnet_modes}
  \vspace{2pt}
  \begin{tabular}{lcccc}
    \toprule
    \multirow{2}{*}{System} &
    \multicolumn{2}{c}{Tioga} &
    \multicolumn{2}{c}{Maple} \\
    \cmidrule(lr){2-3} \cmidrule(lr){4-5}
    & ReqReq & ReqRep & ReqReq & ReqRep \\
    \midrule
    GICC      & \textbf{4.01} & \textbf{3.58} & {11.76} & {11.71} \\
    GASNet-EX & 4.36        & 4.01        & \textbf{3.45}        & \textbf{3.28}        \\
    \bottomrule
  \end{tabular}
\end{table}

Table~\ref{tab:short_am_gasnet_modes} reports the AM microbenchmark results on Tioga and Maple. The evaluation uses GASNet-EX's official \texttt{testam} program and focuses on \emph{short} messages. This choice is intentional: short AMs carry no intrinsic payload/workload and primarily serve as a control-path primitive that triggers the receiver-side handler, making them well-suited for isolating pure messaging and dispatch overhead. Two communication semantics are considered. ReqReq measures a symmetric request--request fast path, reflecting scenarios where both endpoints actively initiate communication, while ReqRep captures a canonical RPC-style request--reply pattern. Comparing the two helps diagnose whether the reply path is optimized differently or incurs additional overhead.

On Tioga, GICC achieves lower AM latency than GASNet-EX in both modes, consistent with its more efficient communication path. On Maple, however, the GPU-triggered variant of GICC exhibits noticeably higher latency than GASNet-EX. This behavior aligns with NVIDIA's guidance for InfiniBand/RDMA environments~\cite{nvidia_nvshmem_performance_2025}: GPU threads generally construct and post work-queue entries less efficiently than CPU threads, which increases the per-message latency when issuing individual operations from the device. Consequently, in a single-AM setting with minimal computation, CPU-driven GASNet-EX can outperform GPU-triggered GICC. Importantly, this microbenchmark intentionally excludes the GPU-kernel launch/termination overhead that is often unavoidable in real coordination logic. If an AM handler ultimately drives GPU-side computation (e.g., launching a kernel or synchronizing GPU work), the host-side orchestration and kernel launch overhead on the GASNet-EX path may dominate and offset the latency advantage suggested by Table~\ref{tab:short_am_gasnet_modes}.

\begin{figure}[t]
    \centering
    \includegraphics[width=.9\linewidth]{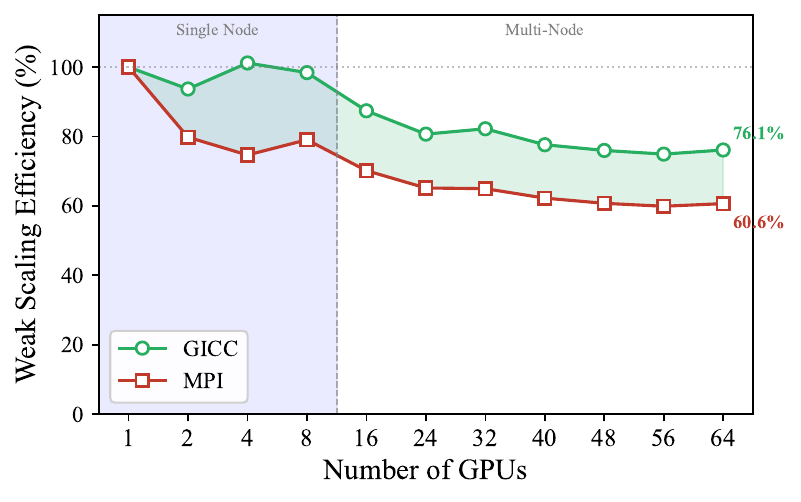}
    \caption{Weak-scaling efficiency of a 2D Jacobi stencil on Tioga (AMD MI250X + HPE Slingshot).}
  \label{fig:jacobi}
\end{figure}

\subsection{Weak-scaling: Jacobi}

We evaluate weak-scaling performance using a 2D Jacobi stencil solver developed by NVIDIA~\cite{nvidia_multigpu_jacobi}, also evaluated in recent NVSHMEM weak-scaling work~\cite{sc25w}. Each GPU maintains a constant local domain of $2048 \times 256$ grid points as the number of GPUs increases, exchanging halo regions with its neighboring ranks at each iteration. We define weak-scaling efficiency as $E_p = T_1 / T_p$, where $T_1$ is the runtime for the single-GPU baseline at the same per-GPU problem size and $T_p$ is the runtime on $p$ GPUs. Figure~\ref{fig:jacobi} shows the weak-scaling efficiency on Tioga. Within a single node (1--8 GPUs), GICC sustains over 93\% efficiency by leveraging GPU IPC for direct peer-to-peer transfers, whereas MPI efficiency drops to approximately 75--80\% due to protocol overhead. As execution scales across multiple nodes, GICC continues to outperform MPI: at 64 GPUs, GICC maintains 76.1\% weak-scaling efficiency compared to MPI's 60.6\%. This gap reflects the combined benefits of intra-node IPC and inter-node GPU-triggered communication.

\subsection{Strong-scaling: Matrix Multiplication}

We next evaluate a ring-exchange communication pattern using an application that implements the Cannon algorithm to compute the square matrix product \( C = A \times B \)~\cite{diomp}. In the experiment, the application uses an additional block stripe for matrix \( B \) to enable overlap of computation and communication. Specifically, we set the number of processes (GPUs) as \( P \), the matrix size as \( N \), and the block stripe width as \( N_s = N / P \). During execution, each process (rank) completed \( P \) compute stages, and each stage involved one communication step in the ring plus a workload of \( N \cdot N_s \cdot N_s \).

\begin{figure}[t]
    \centering
    \includegraphics[width=\linewidth]{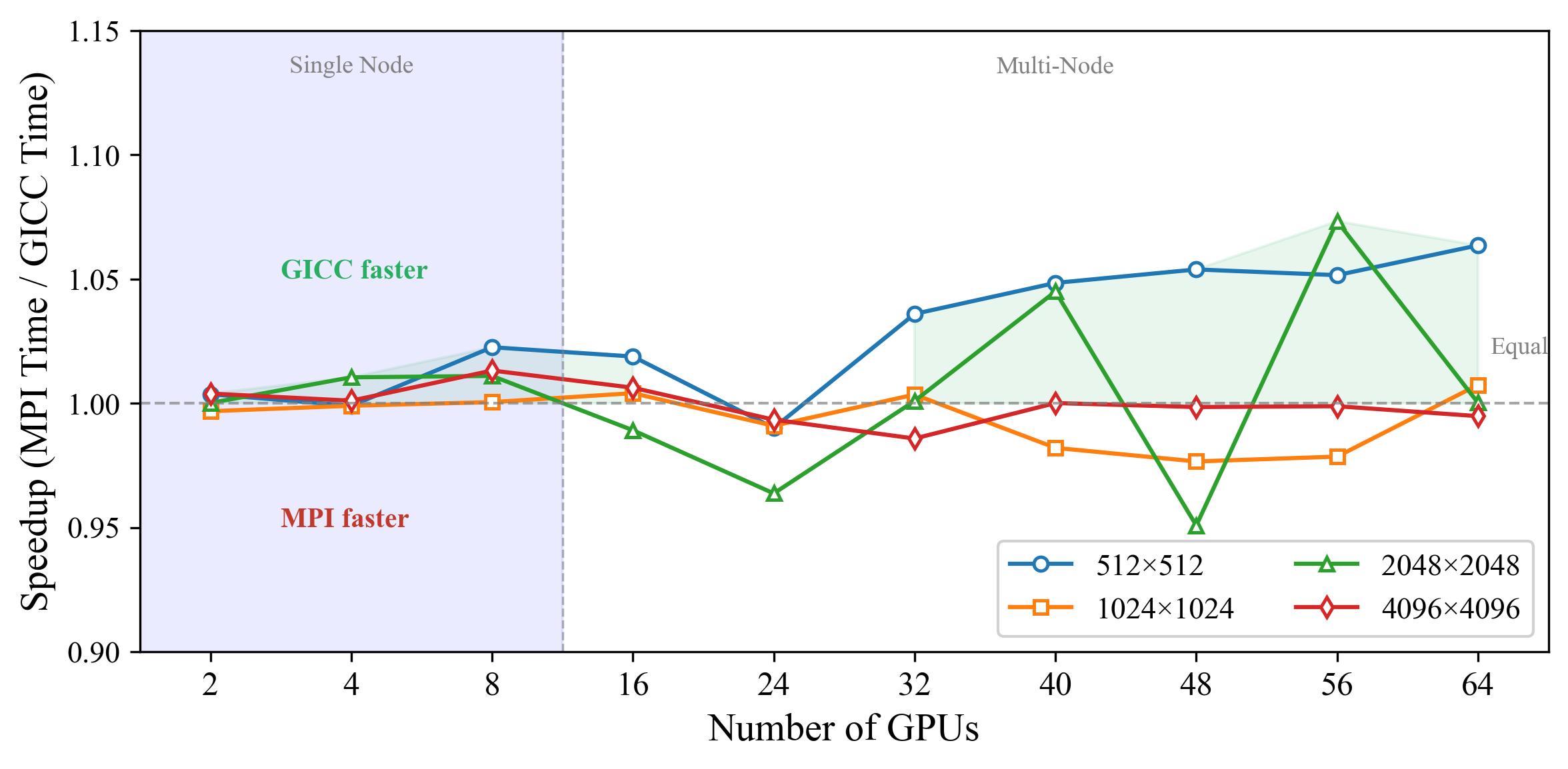}
    \caption{Speedup of GICC over MPI for distributed matrix multiplication across four matrix sizes (512--4096). Speedup $>1$ indicates GICC outperforms MPI.}
  \label{fig:mm}
\end{figure}

Figure~\ref{fig:mm} shows the speedup of GICC over MPI for distributed matrix multiplication on Tioga. For smaller matrices (e.g., 512$\times$512), which involve smaller message sizes per transfer, GICC outperforms MPI by up to 6.4\%. For larger matrices (1024--4096), the two implementations are often close, and several cases modestly favor MPI. This result is consistent with the communication structure of Cannon's algorithm: each stage performs only one ring communication per rank, so there are fewer opportunities to amortize coordination overhead than in Jacobi or Minimod, where each iteration contains repeated fine-grained exchanges. In other words, GICC provides the largest benefit in phase-heavy workloads with frequent coordination, whereas single-exchange-per-stage patterns expose less coordination overhead to eliminate.

\vspace{-8pt}
\subsection{Minimod}

\begin{figure*}[t]
    \centering
    \includegraphics[width=.82\linewidth]{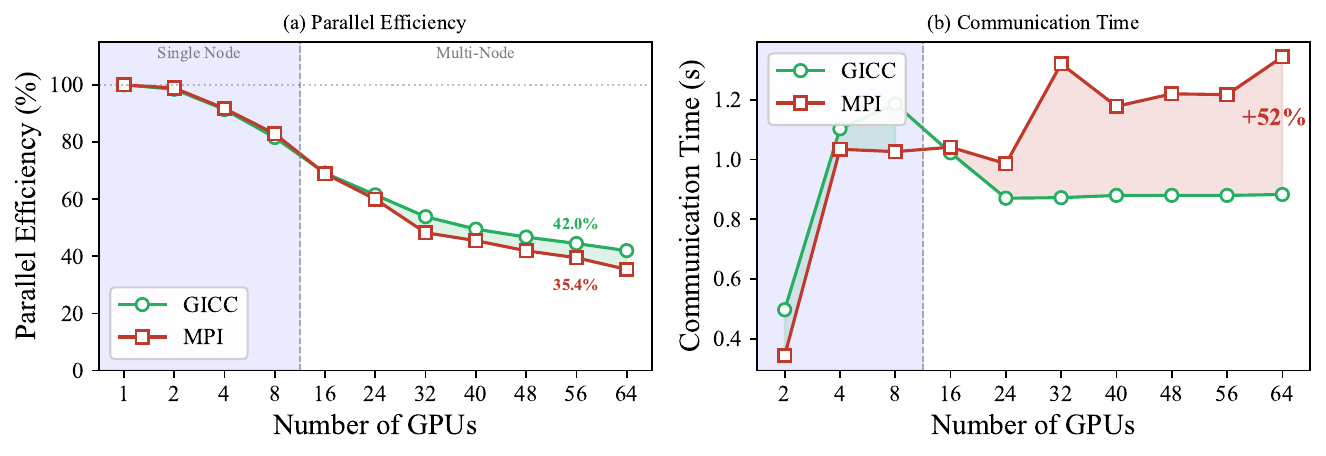}
    \caption{Performance comparison of GICC and MPI implementations for Minimod on Tioga (AMD MI250X + HPE Slingshot): (a) parallel efficiency and (b) communication time.}
  \label{fig:minimod_ofi}
\end{figure*}


Minimod~\cite{meng2020minimod} is an industry-developed proxy application derived from a large-scale seismic simulation code, capturing representative high-order finite-difference stencil computations; stencil workloads are widely used for evaluating GPU programming models and communication performance~\cite{shan2024stencileval}. This study focuses on its acoustic isotropic kernel with a problem size of $1000^3$ grid points, chosen to ensure sufficient computational workload per GPU and realistic communication-to-computation ratios at scale. We use a multi-kernel configuration in which the inner stencil computation, source injection, and boundary condition handling are separate kernels to expose repeated inter-kernel coordination. Under this decomposition, each rank exchanges boundary data with its geometric neighbors after the relevant kernel phases.

Figure~\ref{fig:minimod_ofi} shows the parallel-efficiency and communication-time comparison between GICC and MPI implementations across 1 to 64 GPUs. Within a single node, both implementations exhibit comparable efficiency, achieving over 80\% at 8 GPUs. However, as execution scales beyond a single node, GICC demonstrates stronger scalability, maintaining 42.0\% parallel efficiency at 64 GPUs compared to 35.4\% for MPI. The communication-time analysis in Figure~\ref{fig:minimod_ofi}(b) reveals the underlying cause: the multi-kernel structure requires frequent inter-kernel coordination, and MPI's host-driven approach incurs growing overhead at each kernel boundary. GICC instead allows each kernel to trigger communication directly from the GPU, overlapping it with concurrent kernel execution without host intervention. At 64 GPUs, MPI incurs 52\% higher communication time than GICC, confirming that GPU-driven coordination is particularly advantageous for multi-kernel decompositions.

\section{Related Work}

Existing GPU communication stacks largely remain host-centric: MPI libraries and traditional GPU-aware collectives orchestrate network operations on the CPU, even when data movement bypasses host memory via GPUDirect RDMA \cite{mpi40, potluri2013efficient, nccl}. MPI-based remote GPU offloading frameworks further exemplify this host-centric pattern, where MPI orchestrates device kernels and data movement from CPU-hosted runtime layers \cite{shan2023mpiromp, lu2022efficient}. This control-path placement introduces host--device synchronization and kernel boundary overheads that can serialize distributed control flow \cite{oden2014gpucontrolled}. Some MPI implementations, such as Cray MPICH, introduce stream-triggered communication, but such mechanisms are initiated by stream operations rather than directly by GPU kernel threads \cite{hpecray_mpi_gpunic_async, namashivayam2025gpucentric}.

Device-side communication models such as NVSHMEM and rocSHMEM provide PGAS-style \texttt{put}/\texttt{get} operations from GPU threads \cite{nvshmem}. Beyond direct device-side libraries, PGAS abstractions have also been explored at the programming-model layer for distributed GPU execution \cite{shan2024pgasomp}. However, on OFI-based fabrics these libraries still rely on host-driven progress and therefore remain host-mediated in practice \cite{sc25-gpuinit}. GIN \cite{gin} provides GPU-initiated network operations but, like NVSHMEM, only targets InfiniBand-class environments. Cray MPICH also advertises a kernel-triggered mode in which communication can be launched by GPU threads, but that path does not currently expose the coordination primitives studied here, such as barriers and active messages; moreover, we were unable to enable it on our target Slingshot systems (Section~\ref{sec:evaluation}).
Research systems have explored deeper device/network integration---including GPU-side networking abstractions (GPUnet) and in-network handler execution (sPIN) \cite{kim2014gpunet, hoefler2017spin}---and hardware interfaces for triggered execution (Portals 4, libfabric CXI triggered operations) \cite{brightwell2012portals, libfabric, libfabric-cxi-docs}. These mechanisms expose powerful building blocks, but they are constrained by finite NIC-resident state; prior offload designs often either assume ample resources or require heavyweight reclamation \cite{dosanjh2017offloading}. Complementary work at the programming-model layer has explored device-to-device collective communication in OpenMP target offloading \cite{shan2025devcol} and portable distributed heterogeneous OpenMP runtimes \cite{diomp}, which share GICC's motivation to keep GPU execution at the center of distributed computation. GICC builds on these trends but targets a different missing capability: portable, \emph{sustained} GPU-driven coordination under tight NIC resource limits; it keeps the host off the fast path while ensuring safe resource reuse.

\section{Discussion}

\noindent\textbf{Current Limitations:} While GICC enables GPU-driven coordination on OFI-based fabrics, fundamental constraints of the OFI execution model impose limitations that affect both performance and programmability. On OFI/CXI, GPUs cannot truly \emph{initiate} network operations---they can only \emph{trigger} work that has been pre-armed by the host. This distinction has two consequences. First, the set of possible coordination actions must be bounded and prepared in advance, so the current design is best suited to regular or semi-regular communication patterns rather than fully dynamic communication that emerges at runtime. Second, the finite NIC resources documented in Section~\ref{sec:problem} impose hard ceilings on the number of concurrent coordination instances, forcing GICC to employ sliding-window pipelining and epoch-based reclamation that can introduce back-pressure when coordination frequency exceeds the host's re-arming throughput.

\noindent\textbf{Programming Model Portability:} The current GICC interface exposes a unified API across InfiniBand and OFI fabrics, but the underlying execution semantics differ: on InfiniBand, GPU threads directly post and complete network operations, whereas on OFI, they trigger pre-staged work. This semantic gap complicates portable reasoning about resource usage, ordering, and progress guarantees. Applications tuned for one fabric may exhibit unexpected behavior on another---for example, code that issues an unbounded number of coordination requests will run correctly on InfiniBand but may stall on OFI due to resource exhaustion.

\noindent\textbf{Future Directions:} We plan to extend GICC to additional cloud and HPC fabrics, notably the AWS Elastic Fabric Adapter (EFA), which exposes a similar triggered-operation model via its SRD transport and could benefit from GICC's resource-aware coordination design. Although implementing a full-featured collective communication library is beyond the scope of this work---production systems such as NCCL employ a wide array of sophisticated algorithmic and transport-level optimizations---the techniques developed for our barrier implementation, namely efficient multiplexing of finite NIC resources and lightweight host-side progress, establish a reusable foundation for realizing broader GPU-triggered collectives on resource-constrained fabrics. More broadly, we envision compiler support that lets applications express coordination intent in a fabric-agnostic manner while automatically lowering primitives to fabric-specific backends---direct NIC posting on InfiniBand, triggered DWQ sequences on OFI/CXI, or hybrid paths on emerging interconnects---thereby improving both portability and performance without sacrificing backend-specific optimizations.

\section{Conclusion}

This paper presents GICC, a GPU-driven communication and coordination framework 
for modern HPC systems. GICC addresses two complementary gaps in existing GPU 
communication frameworks. First, GICC provides an efficient GPU-triggered 
communication layer across both InfiniBand and OFI-based fabrics such as HPE Slingshot, 
with the primary design novelty targeting OFI systems where existing runtimes such as NVSHMEM remain host-mediated. On OFI, GICC enables 
GPU-triggered data movement with threshold-based batching that reduces 
triggering overhead from $O(n)$ to $O(1)$ for $n$-operation sequences. Second, 
built on this communication layer, GICC introduces GPU-driven coordination 
primitives---including active messages and synchronization---that allow GPU control flow to 
govern distributed synchronization without host involvement on the fast path. 
To sustain GPU-driven coordination under finite NIC resources, GICC employs a 
hybrid progress model with sliding-window pipelining and epoch-based resource 
reclamation, keeping the host off the critical path while ensuring safe reuse of 
bounded NIC state. Our evaluation on NVIDIA and AMD GPUs over both InfiniBand and
Slingshot demonstrates up to $229\times$ reduction in coordination overhead and up to $1.95\times$ lower put latency than NVSHMEM. In real applications, GPU-aware MPI incurs over 52\% higher communication time than GICC. More broadly, the results show that GICC is most beneficial for phase-heavy workloads with frequent coordination, while single-exchange communication patterns inherently offer less coordination overhead for GICC to eliminate.

\begin{acks}
This material is based upon work supported by the National Science Foundation under Grant No. CCF-2113996.
We would like to thank TotalEnergies E\&P Research and Technologies US for their support of this work.
This research used resources of the Argonne Leadership Computing Facility, which is a U.S. Department of Energy Office of Science User Facility operated under contract DE-AC02-06CH11357.
\end{acks}

\bibliographystyle{ACM-Reference-Format}
\bibliography{software}


\end{document}